\documentclass{JAC2003}
\usepackage{graphicx}
\usepackage{booktabs}

\begin{document}
\title{
TWO CELL REPETITIVE ACHROMATS AND FOUR CELL 
ACHROMATS BASED ON MIRROR SYMMETRY
\vspace{-0.5cm}}
\author{V.Balandin\thanks{vladimir.balandin@desy.de}, 
R.Brinkmann, W.Decking, N.Golubeva \\
DESY, Hamburg, Germany}

\maketitle

\begin{abstract}
In this article we introduce a group-theoretical point of view 
for the design of magnetic optical achromats based on symmetry. 
As examples we use two-cell repetitive achromats and four-cell 
achromats employing mirror symmetry.
\end{abstract}

\section{INTRODUCTION}

As an achromat we will understand a particle transport system
whose linear transfer matrix is dispersion free (first-order achromat)
and whose transfer map does not have nonlinearities of transverse 
motion up to a certain order $n$ ($n$-order achromat).
Although first-order achromats are known and are widely used since the mid 
of 1950s~\cite{Panovsky}, the possibility to use second-order achromats 
in practical accelerator designs was considered
as unrealistic for a long time. It is related to the fact that even with
mid-plane symmetry taken into account the transfer map of a 
magnetostatic system can have as much as 18 independent 
transverse second-order aberrations
and thus requires at least the same number of independent sextupoles 
(or sextupole families) for their correction.
So it was somewhat a surprise when at the end of 1970s
the design of a four-cell second-order achromat which uses
only 8 sextupoles arranged in only two independent families
was found~\cite{Brown}. The theory developed in that paper
(theory of achromats based on repetitive symmetry)
states, that any system built out of $n$ identical cells ($n > 1$)
with the overall transfer matrix equal to the identity matrix (except, possibly, 
for the element relating the time-of-flight difference to the energy 
difference) and with the tunes of a cell such that resonances which
are not forbidden by the mid-plane symmetry are avoided to third order
gives a first-order achromat and can be corrected to become a
second-order achromat using only two families of sextupoles.  
Later on the theory of achromats, which employ mirror (reversal) symmetry, 
was also developed in~\cite{WanBerz}. 

Even if it is clear that automatic cancellation of some aberrations
in the symmetry based achromat designs follows from the symmetry of the magnet 
arrangement in the system, there are several questions which are not yet 
answered by any of the presently known theories.
What is the exact role of the symmetry of the magnet arrangement?
Why is the system transfer matrix
equal to the unit matrix in both transverse planes
in almost all known  achromats (with the exception of a magnifying magnetic 
optical achromat~\cite{BrownServranckx})?
And the most important question is, is there a magnet arrangement
which will give better cancellation of aberrations than those already known?

In this paper we will show a way to answer these questions by looking at the  
symmetry based achromat designs from the point of view of the theory
of finite matrix groups.
In our approach the system is a good candidate for making a second-order 
achromat if its aberrations, when represented in the form of a third order
homogeneous polynomial entering the Lie exponential factorization of the system 
transfer map, form not an arbitrary polynomial but a polynomial which is an
invariant under the action of some finite matrix group.
As long as the transverse particle motion is uncoupled and
as long as any finite subgroup of the group of two by two invertible matrices
is either $C_n$ (cyclic group of order $n$) or 
$D_n$ (dihedral group of order $2 n$), from the point of view of our theory 
there is no better cancellation of aberrations than  
provided by the action of $C_n$ or $D_n$ groups.
Nevertheless, in our opinion it does not mean that one has to stop looking for
new appearances of cyclic and dihedral groups in the magnetic systems.
For example, the cyclic group shows itself not only in the repetitive achromats
constructed from identical cells,
but also in staircase achromats, in magnifying achromats, and probably 
some other useful manifestations of this group could be found. 

Due to space limitation, in this paper we will consider
only two-cell second-order repetitive achromats (group $C_2$) and
four-cell second-order achromats based on mirror symmetry (group $D_2$).
The general theory including a discussion of higher order achromats
and invention of achromats based on arbitrary order dihedral group
will be presented in a separate publication~\cite{GroupsAndAchr}.
We have selected these two achromats not 
only because they are manifestations of the smallest nontrivial cyclic group
and of the smallest dihedral group which is not a cyclic group, but also
because they were achromats which actually motivated this investigation and which
we studied first during the design of the arcs of the
post-linac collimation section of the European XFEL Facility~\cite{XFEL,ColXFEL}.

\section{MAPS AND ACHROMATS}

We will consider the beam dynamics in a mid-plane symmetric
magnetostatic system and will use a complete set of symplectic variables
$\mbox{\boldmath $z$} = (x, p_x, y, p_y, \sigma, \varepsilon)$
as particle coordinates. 
In this set the variables $\hat{\mbox{\boldmath $z$}} = (x, p_x, y, p_y)$
describe the transverse particle motion and 
the variables $\sigma$ and $\varepsilon$ characterize
the longitudinal dynamics.
We will represent particle transport from the longitudinal
location $\tau_1$ to the location $\tau_2$ by a symplectic map ${\cal M}_{1,2}$ and 
we will assume that for arbitrary positions $\tau_1, \tau_2$
the point $\mbox{\boldmath $z$} = \mbox{\boldmath $0$}$ is the fixed
point and that the map ${\cal M}_{1,2}$ can be Taylor 
expanded in its neighborhood.
If a beam passes successively through the maps 
${\cal M}_{1,2}$ and ${\cal M}_{2,3}$
then we will use the following notation for the map of 
the composite system\footnote{Since the chance of confusion is
small, we will use symbol $:*:$ to denote both, map compositions
and exponential Lie operators.} 
 
\noindent
\begin{eqnarray}
:{\cal M}_{1,3}: \,=\,
:{\cal M}_{2,3}({\cal M}_{1,2}): \,=\,
:{\cal M}_{1,2}: \,:{\cal M}_{2,3}: .
\label{MAA_2}
\end{eqnarray}

Up to any predefined order $n$ the aberrations of a map $\,{\cal M}\,$
can be represented through a Lie factorization as 

\noindent
\begin{eqnarray}
:{\cal M}: \,=_n\,
\exp(:{\cal F}_{n + 1} + \ldots + {\cal F}_3:) :M: ,
\label{IFB_3}
\end{eqnarray}

\noindent
where each of the functions ${\cal F}_m$
is a homogeneous polynomial of degree $m$ in the variables $\mbox{\boldmath $z$}$
and the symbol $=_n$ denotes equality up to order $n$ 
when maps on both sides of (\ref{IFB_3})
are applied to the phase space vector $\mbox{\boldmath $z$}$.
Using this representation we will say that the map $\,{\cal M}\,$ 
is a $n$-order achromat if the matrix of its linear part
$M$ is dispersion free and all polynomials 
${\cal F}_m$ are functions of the variable
$\varepsilon$ only.

We will use that the map ${\cal M}$ of a magnetic system  which is 
symmetric about the horizontal midplane $\,y = 0\,$ satisfies

\noindent   
\begin{eqnarray}
:{\cal M}:\,:T_M: \,=\,: T_M: \,:{\cal M}: ,
\label{TWO_C2_6}
\end{eqnarray}

\noindent   
where $T_M = \mbox{diag}(1, 1, -1, -1, 1, 1)$
is the mid-plane symmetry matrix.

\section{TWO CELL REPETITIVE ACHROMAT AND CYCLIC
GROUP OF ORDER TWO}

Let us consider a system constructed by a repetition of 
two identical cells with the cell map ${\cal M}_c$ given by
the following Lie factorization

\noindent
\begin{eqnarray}
:{\cal M}_c: \,=_2\,
\exp(:{\cal F}_3(\mbox{\boldmath $z$}):) :M_c:.
\label{TWO_C_1}
\end{eqnarray}

\subsection{First-Order Conditions}

The necessary and sufficient conditions for a lattice made of 
two identical cells to be a first-order achromat are 

\noindent
\begin{eqnarray}
\left(
\begin{array}{cc}
r_{11} & r_{12}\\
r_{21} & r_{22}
\end{array}
\right)
\left(
\begin{array}{c}
r_{16}\\
r_{26}
\end{array}
\right)
= -
\left(
\begin{array}{c}
r_{16}\\
r_{26}
\end{array}
\right),
\label{TWO_C_2}
\end{eqnarray}

\noindent
where $\,r_{mk}\,$ are the elements of the cell matrix $\,M_c$.

\subsection{Second-Order Conditions}

Let us assume that the linear achromat 
conditions (\ref{TWO_C_2}) are satisfied.
Then it is possible to show that the equation

\noindent   
\begin{eqnarray}
\left(
\begin{array}{cc}
r_{11} & r_{12}\\
r_{21} & r_{22}
\end{array}
\right)
\left(
\begin{array}{c}
A\\
B
\end{array}
\right)
+
\left(
\begin{array}{c}
r_{16}\\
r_{26}
\end{array}
\right)
=
\left(
\begin{array}{c}
A\\
B
\end{array}
\right)
\label{ABC435}
\end{eqnarray}
 
\noindent   
can always be solved with respect to the variables $A$ and $B$.
And though in some rare cases the solution can be non-unique, one
sees that any $A$ and $B$ satisfying (\ref{ABC435}) give
initial conditions for the cell periodic coordinate and
momentum dispersion functions.
This means that if condition (\ref{TWO_C_2}) is satisfied, 
the matrix $\,M_c\,$ can be written
in the form

\noindent
\begin{eqnarray}
M_c\,=\,D_c\, N_c\, D^{-1}_c ,
\label{TWO_C_3}
\end{eqnarray}

\noindent
where the matrix $N_c$ is dispersion-free
and the matrix $D_c$
can be represented in the form of a Lie operator as follows

\noindent   
\begin{eqnarray}
:D_c: \,=\, \exp(:\varepsilon \, (B \, x  \,-\, A \, p_x):).
\label{TWO_C_4}
\end{eqnarray}

\noindent   
Using (\ref{TWO_C_3}) the cell transfer map
can be brought into the form

\noindent
\begin{eqnarray}
:{\cal M}_c: \,=_2\,
:D_c:^{-1}
\exp(:{\cal P}_3(\mbox{\boldmath $z$}):) 
\,:N_c:\, :D_c:
\label{TWO_C2_1}
\end{eqnarray}

\noindent
with $\,{\cal P}_3(\mbox{\boldmath $z$})=
{\cal F}_3(x + A \varepsilon, 
p_x + B \varepsilon, y, p_y, \varepsilon)$,
and for the map of the two cell system 
$:{\cal M}:=:{\cal M}_c::{\cal M}_c:$
we obtain

\noindent
\begin{eqnarray}
:{\cal M}:\,=_2\,
:D_c^{-1}:
\exp(:2 \cdot {\cal S}_3(\mbox{\boldmath $z$}):)
\,:D_c: \,:M:,
\label{TWO_C2_3}
\end{eqnarray}

\noindent
where $M = M_c M_c$ is the system transfer matrix,

\noindent
\begin{eqnarray}
{\cal S}_3(\mbox{\boldmath $z$}) \,=\,
(1\,/\,2) \cdot
\big(
{\cal P}_3(\hat{\mbox{\boldmath $z$}}, \,\varepsilon)
\,+\,
{\cal P}_3(M_4 \,\hat{\mbox{\boldmath $z$}}, \,\varepsilon)
\big),
\label{TWO_C2_3_0}
\end{eqnarray}

\noindent
and the four by four matrix $M_4$ is the upper left block
of the six by six matrix $N_c$
(or, equivalently, of the matrix $M_c$).
Thus, according to the representation (\ref{TWO_C2_3}), 
the two cell first-order achromat will become a 
second-order achromat if, and only if, the function
${\cal S}_3$ will be a function of the variable 
$\varepsilon$ only, i.e. if
$\,{\cal S}_3(\hat{\mbox{\boldmath $z$}}, \,\varepsilon)
\,-\,
{\cal S}_3(\mbox{\boldmath $0$}, \,\varepsilon)
\,=\,0$.

\subsection{Appearance of the Cyclic Group $\,C_2\,$\\
and Role of the Mid-Plane Symmetry}

For the mid-plane symmetric system the polynomial ${\cal F}_3$ 
can have as much as 18 nonzero monomials responsible for the independent 
transverse aberrations.
Why should one expect that the polynomial ${\cal S}_3$ has
a smaller number of them, i.e. why should one expect that 
the map of the two cell system has less 
independent second order aberrations than the cell map? 
No reason is seen for that
in the case of an arbitrary matrix $M_4$.
The situation will change, if we assume that
$M_4^2 = I_4$ while $M_4 \neq I_4$, where
$I_m$ is the $m$ by $m$ identity matrix.
With this assumption the matrices $I_4$ and $M_4$ will form
a finite matrix group, which is isomorphic to the group $C_2$,
and ${\cal S}_3$ will not be an arbitrary polynomial
anymore. It becomes 
the result of the application of the group Reynolds
(averaging) operator to the polynomial ${\cal P}_3$ and
for an arbitrary ${\cal P}_3$ is a 
polynomial which remains invariant under the group action.
As an abstract object the group $C_2$ is unique,
but there are three different
choices for the matrix $M_4$ in order to satisfy the group condition

\noindent
\begin{eqnarray}
M_4 = \left\{
\begin{array}{l}
\mbox{diag}(-I_2, -I_2)\\
\mbox{diag}(\;\;\;I_2, -I_2)\\
\mbox{diag}(-I_2, \;\;\; I_2)
\end{array}
\right.
.
\label{TWO_C2_5_1}
\end{eqnarray}

\noindent
Before considering the optimal choice for the matrix 
$M_4$ from the list (\ref{TWO_C2_5_1}), 
let us discuss shortly the role of the mid-plane symmetry.
The commutation relation (\ref{TWO_C2_6}) tells us
that $\,{\cal F}_3\,$ in (\ref{TWO_C_1}) is not an
arbitrary polynomial, but is an invariant of another
$C_2$ group (mid-plane symmetry group) formed by the matrices 
$\,I_6\,$ and $\,T_M$. So as a total symmetry group of the 
mid-plane symmetric two-cell system  
one can consider the group generated by the matrix $\,M_4\,$ and by
the four by four upper left block of the matrix $T_M$,
and then study the action of the Reynolds operator of this
group on arbitrary polynomials.
Or, as we prefer, one can use a slightly different approach
and utilize the fact that the Reynolds operator 
of the group formed by the matrix $M_4$
maps the set of invariants of the mid-plane symmetry group into itself. 

With all these discussions one can show that 
as optimal choice for the matrix $\,M_4\,$
one can take either the first or the third line in (\ref{TWO_C2_5_1}).
Both give the number of the remaining independent transverse 
aberration to be corrected by sextupole magnets equal to six.
With both these choices the condition (\ref{TWO_C_2}) will be
satisfied automatically for an arbitrary cell dispersions $r_{16}$
and $r_{26}$, and the transfer matrix of the two cell system will
be equal to the identity matrix in both transverse planes.

\section{FOUR CELL ACHROMAT BASED ON REFLECTION SYMMETRY
AND THE KLEIN FOUR-GROUP}

Following~\cite{WanBerz} let us consider a system where the
forward cell is followed by a reversed cell and then this two-cell
configuration is repeated once more, i.e. let us consider a system
whose transfer map is given by the following relation

\noindent
\begin{eqnarray}
:{\cal M}: = :{\cal M}_F: :{\cal M}_R: :{\cal M}_F: :{\cal M}_R:.
\label{FOUR_1}
\end{eqnarray}

\noindent
It is clear that 
with $:{\cal M}_c: = :{\cal M}_F: :{\cal M}_R:$
this system can be treated as the two-cell
system considered in the previous section
and, therefore, if the first-order achromat conditions (\ref{TWO_C_2})
are satisfied, then the matrix of the half of the
system $M_c = M_R M_F$ can be represented in the form (\ref{TWO_C_3}).
The property that now the cell ${\cal M}_c$ is mirror symmetric
with respect to its center allows us to prove additionally that $B$
entering formula (\ref{TWO_C_4}) is equal to zero 
and therefore the matrix $D_c$ commutes with the reversal symmetry
matrix $T_R = \mbox{diag}(1, -1, 1, -1, -1, 1)$.
That is an important fact which not only allows us to obtain
the representation (\ref{FOUR_3}), but also allows to remove 
one superfluous constraint on the linear optics which was
used in~\cite{WanBerz} (see details in~\cite{GroupsAndAchr,ColXFEL}).  

Let us assume that the map of the forward cell ${\cal M}_F$ is 

\noindent
\begin{eqnarray}
:{\cal M}_F: \,=_2\,
\exp(:{\cal F}_3(\mbox{\boldmath $z$}):) :M_F:.
\label{FOUR_2}
\end{eqnarray}

\noindent
Then we can represent the map of the total system as

\noindent
\begin{eqnarray}
:{\cal M}:\,=_2\,
:D_c^{-1}:
\exp(:4 \cdot {\cal S}_3(\mbox{\boldmath $z$}):)
\,:D_c: \,:M:,
\label{FOUR_3}
\end{eqnarray}

\noindent
where $M=M_c M_c$ is the system transfer matrix, 

\noindent
\begin{eqnarray}
{\cal S}_3(\mbox{\boldmath $z$}) \,=\,
(1 \,/\,4)\cdot \sum_{m=1}^{4}
{\cal P}_3(A_m \,\hat{\mbox{\boldmath $z$}}, \,\varepsilon),
\label{FOUR_4}
\end{eqnarray}

\noindent
and the four by four matrices $A_1$, $A_2$, $A_3$ and $A_4$ are the
upper left blocks of the matrices 
$I_6$, $T_R N_c$, $N_c$ and $T_R N_c^2$,
respectively, and
$\,{\cal P}_3(\mbox{\boldmath $z$}) = 
{\cal F}_3(x + A \varepsilon, 
p_x, y, p_y, \varepsilon)$.

Looking at (\ref{FOUR_4}) one sees that if the matrices $A_m$
will form a finite matrix group, then $S_3$ is the result of 
applying the Reynolds operator of this group to the 
polynomial ${\cal P}_3$ and, therefore, is an invariant.
What kind of group could it be?
One can check that while matrices $A_1$ and $A_3$
are symplectic, the matrices $A_2$ and $A_4$ are
antisymplectic. Because $A_1$ is a unit matrix and because
the product of symplectic and antisymplectic matrices is an
antisymplectic matrix, it follows that
in order to have any kind of a group structure 
the matrix $A_3$ should be
equal to its own inverse, i.e. the equation $A_3^2 = I_4$ 
must be satisfied. This equation has four
symplectic solutions $A_3 = \mbox{diag}(\pm I_2, \pm I_2)$ and 
the choice of any of them completely determines the remaining
matrices $A_m$ and also gives us some group structure.
If we take $A_3 = \mbox{diag}(I_2, I_2)$, then
we have two exemplars of the same $C_2$ group and
all other choices will give us the Klein four-group ($D_2$ group).

Similar to the situation considered in the previous section,
there are two optimal choices for the matrix
$A_3$ with the mid-plane symmetry taken into account.
Namely, one can take $A_3 = \mbox{diag}(-I_2, -I_2)$
or $A_3 = \mbox{diag}(-I_2, I_2)$ and one obtains the number
of the independent transverse aberrations left 
to be corrected by sextupoles to be four.

We see that the second-order achromat built by using
the repetition of two identical cells (FF)
requires six independent sextupole families 
while in the second-order achromat constructed from 
two mirror symmetric cells (FRFR)
only four sextupole families are needed.
But because one has to put sextupoles into achromats in
such a way that the symmetry is preserved, the minimum
number of sextupoles is 16 for the symmetry FRFR and
only 12 for the symmetry FF.
Note that this is a general situation in the so-called
``resonant case''. If, in such a case, the number of
independent multipole families is of concern, then it
is better to use achromats based on the $D_n$ group, and if
the total number of multipole magnets has to be minimized,
then achromats utilizing the $C_n$ group perform better
(see details in~\cite{GroupsAndAchr}).

\end{document}